\documentclass[fleqn,12pt,twoside]{article}
\usepackage{espcrc1}

\usepackage{graphicx}

   \newcommand{\nex}{{\sc neXus}}

\title{New results for hadronic collisions in the framework of the Parton-Based Gribov-Regge
Theory}

\author{T. Pierog\address[SUB]{SUBATECH, University of Nantes, France}
	\address{Forschungzentrum Karlsruhe, Institut f\"ur Kernphysik, Karlsruhe,
			      Germany}, 
	H. J. Drescher\address{Physics Department, New York University, 
				New York, USA}, 
	F. Liu\addressmark[SUB]\address{Institute of Particle Physics, 
				Huazhong Normal University, Wuhan, China}, 
	S. Ostaptchenko\address{Institut f\"ur Experimentelle Kernphysik, 
		 	      University of Karlsruhe, Karlsruhe, Germany}
		       \thanks{On leave of absence from Moscow State University, Institute of Nuclear
		               Physics, Moscow, Russia}
	and 
	K. Werner\addressmark[SUB]}
       
\begin{document}

\maketitle

\begin{abstract}
We recently proposed a new approach to high energy nuclear scattering,
which treats hadronic collisions in a sophisticated
way. Demanding theoretical consistency as a minimal requirement for a
realistic model, we provide a solution for the energy conservation,
screening problems and identical elementary interactions, the so-called 
"Parton-Based Gribov-Regge Theory" including enhanced diagrams. 
We can now present some of our results for SPS and RHIC energies.\end{abstract}

\section{INTRODUCTION}

The most sophisticated approach to high energy hadronic interactions is the
so-called Gribov-Regge theory \cite{Gribov:fc}. This is an effective field
theory, which allows multiple interactions to happen ``in parallel'', with
phenomenological objects called \emph{Pomerons} representing elementary interactions
\cite{Baker:cv}. Using the general rules of field theory, one may express cross
sections in terms of a couple of parameters characterizing the Pomeron. 

A big disadvantage of GRT implementations was so far the fact that cross sections 
and particle production are
not calculated consistently: the fact that energy needs to be shared between
many Pomerons in case of multiple scattering is well taken into account when
considering particle production (in particular in Monte Carlo applications),
but not for cross sections \cite{Abramovsky:bw}. 

Another problem is that at high energies, one also needs a consistent
approach to include both soft and hard processes. The latter are usually
treated in the framework of the parton model, which only allows the calculation
of inclusive cross sections. 

We consider now a new approach called Parton-Based Gribov-Regge theory,
in order to solve the above-mentioned problems. We use both the
language of Pomerons (as in Gribov-Regge theory) in order to calculate
probabilities (and related to this: cross sections) and the language
of strings (to treat particle production). But we treat these both
aspects in a consistent fashion. This is the really new and attractive
feature of our approach. The price to be paid are great technical
difficulties which took us a couple of years to solve them. For all
the details and applications of the corresponding Monte Carlo program
\nex~3 see \cite{Drescher:2000ha}.

\section{PARTON-BASED GRIBOV-REGGE THEORY}

\subsection{Basic ideas}

\begin{figure}[htp]
\vskip-0.5cm
\begin{center}
\includegraphics[  scale=0.5]{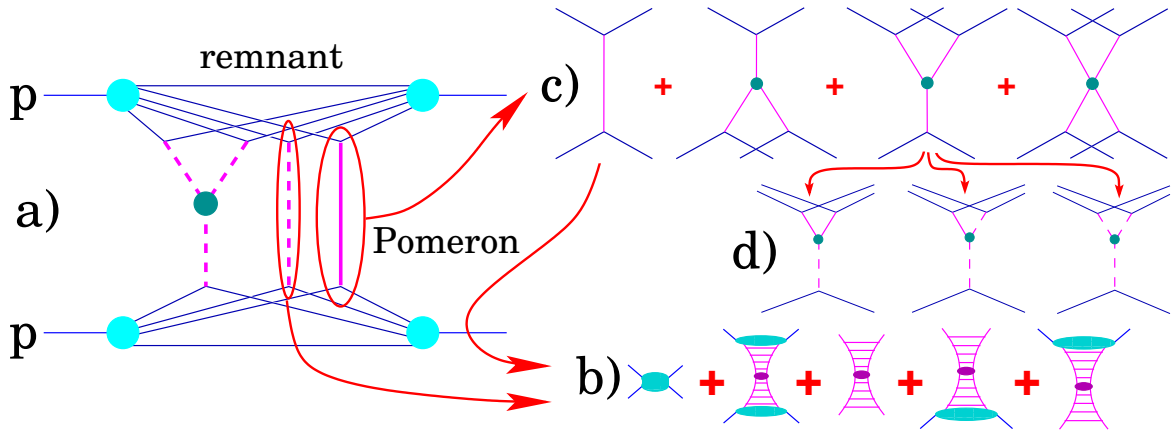}
\end{center}
\vskip-1.5cm
\caption{\label{allin1} a) Multiple elementary interactions (Pomerons)
in \nex. The energy of each proton (blob) is shared between elastic (full
vertical line) and inelastic (dashed vertical line) elementary interactions.
 A Pomeron b) has  soft (blob), hard (ladder)
 and semihard contributions. c) Enhanced diagrams are included and can give
 different inelastic contributions d).}
\vskip-.5cm
\end{figure}

We will discus the basic features of the new approach in a qualitative
fashion. It is an effective theory based on effective elementary interactions.
Multiple interactions happen in parallel in proton-proton collisions.
An elementary interaction is referred to as Pomeron, and can be either elastic
(uncut Pomeron) or inelastic (cut Pomeron). The spectators
of each proton form a remnant, see Fig.~\ref{allin1}a. %

Since a Pomeron is finally identified with two strings, the Pomeron aspect
(to obtain probabilities) and the string aspect
(to obtain particles) are treated in a completely consistent way. \emph{In
both cases energy sharing is considered in a rigorous way, and in
both cases all Pomerons are identical.} 

This theory provides also a consistent treatment for hard and soft
processes: each Pomeron can be expressed in terms of contributions
of different types, soft, hard and semihard, cf. Fig.~\ref{allin1}b.
 A hard Pomeron stands
for a hard interaction between valence quarks of initial hadrons. A semihard
one stands for an interaction between sea quarks but
in which a perturbative process involves in the middle.
No perturbative process occures at all in soft Pomeron.

A Pomeron is an elementary interaction. 
But those Pomerons may interact with each other at high energy \cite{Baker:cv,kai86}, then they give another type of interaction called 
\emph{enhanced diagram}. There are many types of enhanced diagrams depending on 
the number of Pomerons for each vertices and on the number of vertices. 
In our model, 
effective first order of triple and 4-Pomeron vertices (Y and X diagrams see
Fig.~\ref{allin1}c) are enough to cure unitarity problem which occure at high energy
without this kind of diagram \cite{pierog:th}. Indeed, Y-type diagrams are screening
corrections which are negative contributions to the cross-section. X diagram is
anti-screening. The inelastic contributions (cut enhanced diagrams on Fig.~\ref{allin1}d) of this diagrams
contribute to the increase of the fluctuations in particle production.

\subsection{Particle production in \nex}

\begin{figure}[htp]
\vskip-1.cm
{\par \hfill
\begin{center}
\includegraphics[  scale=0.45]{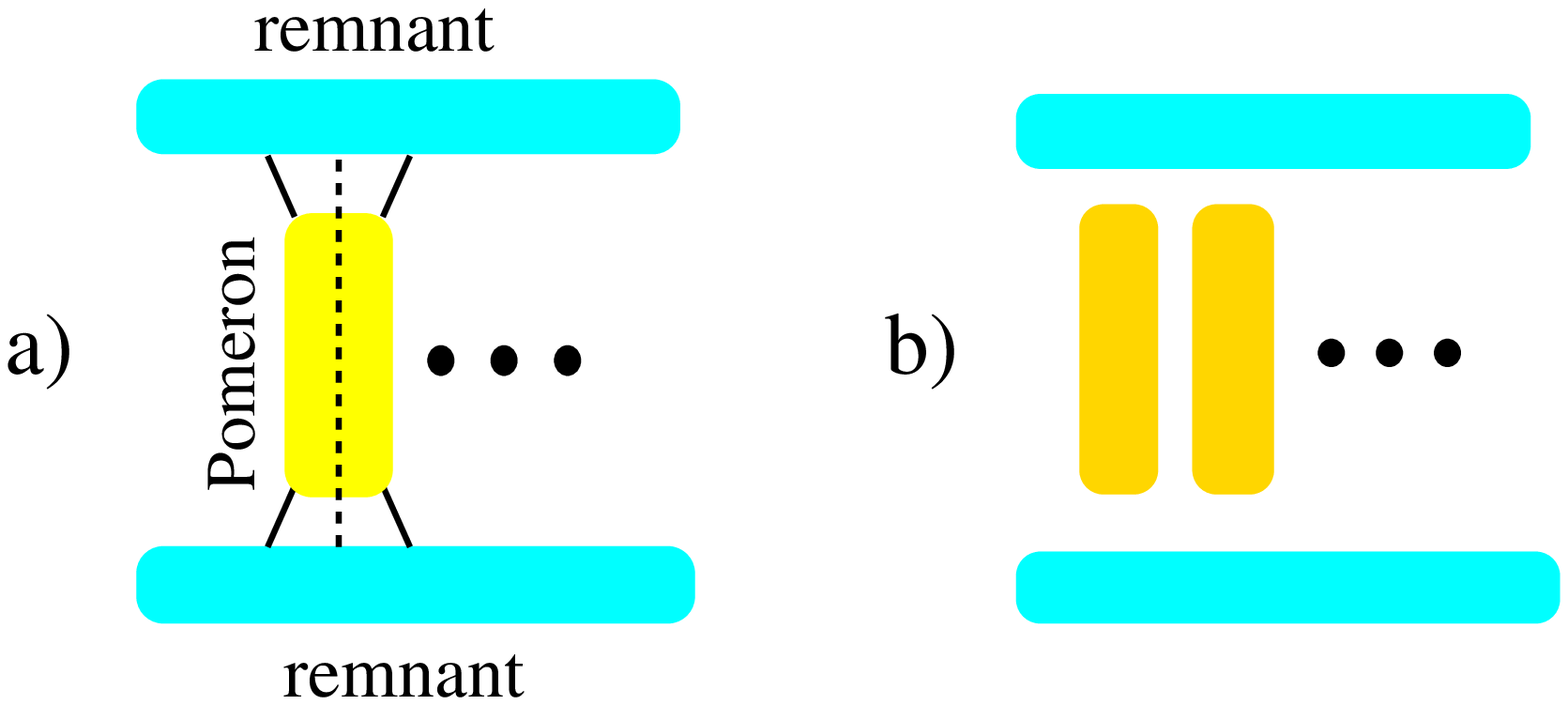}\hspace{1cm}
\includegraphics[  scale=0.5]{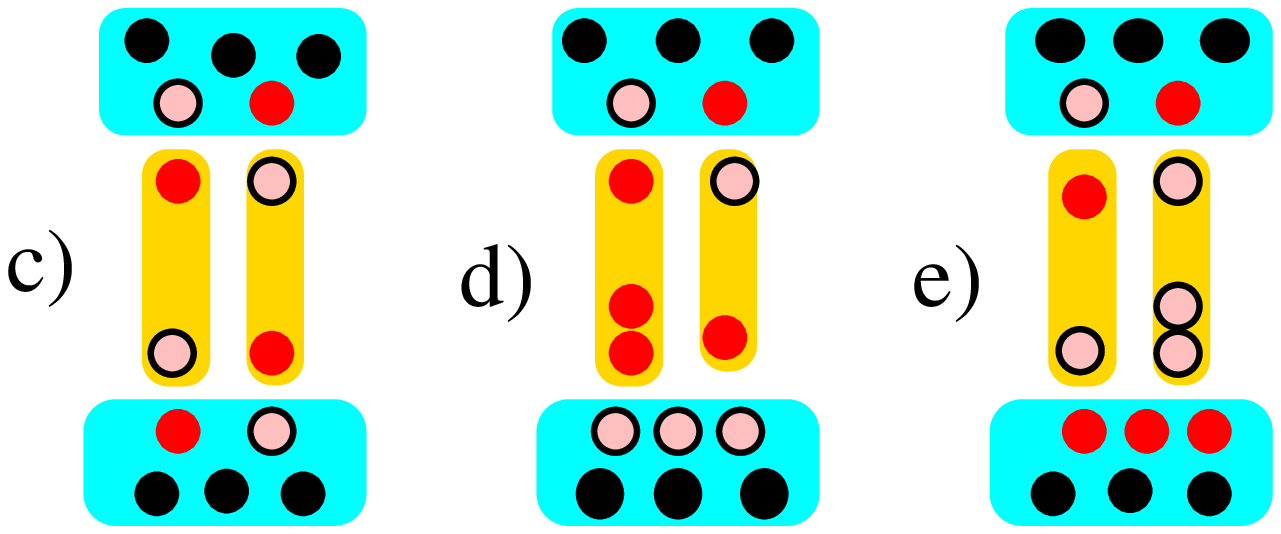}
\end{center}\hfill
\par}
\vskip-1.cm
\caption{ a) Each cut pomeron is regarded as two strings b). 
c) The most simple and frequent collision
configuration has two remnants and only one cut Pomeron represented
by two $\mathrm{q}-\overline{\mathrm{q}}$ strings. d) One of the
$\overline{\mathrm{q}}$ string-ends can be replaced by a $\mathrm{qq}$
string-end. e) With the same probability, one of the $\mathrm{q}$
string-ends can be replaced by a $\overline{\mathrm{q}}\overline{\mathrm{q}}$
string-end. 
\label{nexus1}}
\vskip-.5cm
\end{figure}

Thanks to a Monte Carlo, first the collision configuration is determined:
i.e. the number of each type of Pomerons exchanged between the projectile
and target is fixed and the initial energy is shared between the Pomerons
and the two remnants. Then particle production is accounted from two
kinds of sources, remnant decay and cut Pomeron. A Pomeron may be
regarded as a two-layer (soft) parton ladder attached to projectile
and target remnants through its two legs. Each leg is a color singlet,
of type q$\overline{\mathrm{q}}$ , qqq or 
$\overline{\mathrm{q}}$$\overline{\mathrm{q}}$$\overline{\mathrm{q}}$ from the
sea, 
and then each cut Pomeron is regarded as two strings, cf. Fig.~\ref{nexus1}a. %
It is a natural idea to take quarks and antiquarks from the sea as
string ends for soft Pomeron in \nex, because
an arbitary number of Pomerons may be involved. 
In addition to this soft Pomerons,
hard and semihard Pomerons are treated differently. To give a proper
description of deep inelastic scattering data, hard and some of semihard
Pomerons are connected to the valance quarks of the hadron. 

Thus, besides the three valence quarks,
each remnant has additionally quarks and antiquarks to compensate
the flavours of the string ends, as shown in Fig.~\ref{nexus1}c. 
According to its number of quarks and antiquarks and to the phase space, a
remnant decays into mesons, antibaryons and baryons \cite{Liu:2002gw}. Therefore, 
from remnant decay, baryon production is favored due to the initial 
valence quarks and in particular produce strong leading particle effects 
(two wings in the rapidity spectra) for proton and $\Lambda$.

\section{RESULTS}

\begin{figure}[htp]
\vskip-0.5cm
\begin{center}
\includegraphics[scale=0.65]{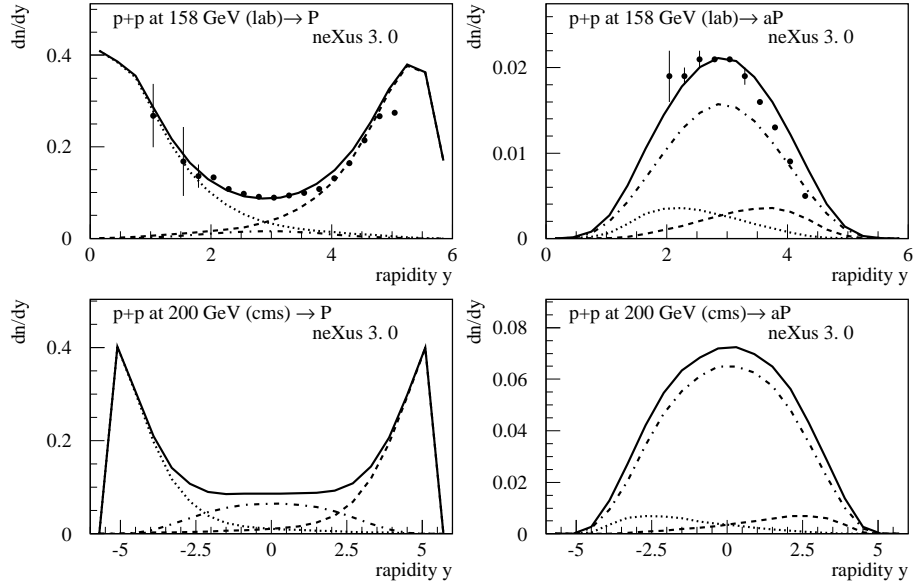}
\end{center}
\vskip-1.5cm
\caption{\label{proton} Proton and anti-proton rapidity spectra for
 NA49 \cite{Kadija:sqm2001} (158 GeV lab) (top) and RHIC (200 GeV cms) (bottom) 
 energies. The remnant contributions are the dashed and the dotted lines, the
 dashed-dotted line is the Pomeron contribution and the full line is the sum.}
\vskip-.5cm
\end{figure}

Considering energy sharing, enhanced diagram and identical elementary
interactions, a great number of particle distributions can be calculated
within \nex~3 for any kind of hadronic and nuclear interaction \cite{pierog:th}. As an example, 
Fig.~\ref{proton} depicts the rapidity spectra for proton
and anti-proton production for 2 different energies. The different contributions for the particle
production (cut Pomeron: dashed-dotted line, remnants: dashed line and dotted
line) show that the assimetry between baryon and  anti-baryon production comes
from the remnant contributions. 

At low energy (158 GeV lab), \nex~is in good agreement with NA49 data~\cite{Kadija:sqm2001}.
For RHIC energy (200 GeV cms), the prediction shows
that the cut Pomeron distribution becomes dominant at midrapidity even for the
proton, which leads to reduce the assymetry in this region (ratio closer to 1).

In conclusion, a new realistic model \nex~solving most of the existing problems in high
energy hadronic interaction models is available \cite{web}.

S. O. was supported by the German Ministry for Education and Research (BMBF).

\end{document}